\documentstyle[twocolumn,aps,epsf]{revtex}

\begin{document}
\draft
\preprint{HEP/123-qed}
\title{Strings with Negative Stiffness and Hyperfine Structure}
\author{M. C. Diamantini\cite{bylinea} and H. Kleinert}
\address{Institut f\"ur Theoretische Physik, Freie Universit\"at Berlin,\\
Arnimallee 14, D-1000 Berlin 33, Germany}
\author{C. A. Trugenberger\cite{bylineb}}
\address{D\'epartement de Physique Th\'eorique, Universit\'e de Gen\`eve,\\
24, quai E. Ansermet, CH-1211 Gen\`eve 4, Switzerland}
\date{\today}
\maketitle
\begin{abstract}
We propose a new string model by adding a higher-order
gradient term to the rigid string, so
that the stiffness can be
positive or negative without loosing stability. In the large-$D$
approximation, the
model has three phases, one of which with a new type of generalized
``antiferromagnetic" orientational correlations. We find an infrared-stable
fixed point describing world-sheets with vanishing tension and Hausdorff
dimension $D_H=2$. Crumpling is prevented by the new term which suppresses
configurations with rapidly changing extrinsic curvature.
\end{abstract}

\narrowtext

{\bf 1.} So far, fundamental strings \cite{gsw} can
be quantized only in critical
spacetime dimensions $D_{\rm c}$, due  to the difficulty of
handling the Liouville field, which does not decouple
in dimensions different from $D_{\rm c}$.
Strings in four space-time dimensions,
however, are of great physical importance
since they are supposed to describe the confining phase of
non-Abelian gauge theories \cite{polchinski1}, as
confirmed by recent
precision numerical tests \cite{caselle}.
As yet, there exists no consistent
quantum theory of such strings.
Attempts in this direction based
on
discrete models for the Liouville field failed
\cite{brezin}, since these could not  be
generalized beyond one dimension. Further
hope was based on
strings with extrinsic curvature stiffness \cite{rigid},
but the
infrared irrelevance of stiffness
did not prevent the world-sheets from crumpling \cite{largen}.

Recent progress in this field is based on a new type of action. In its
local formulation  \cite{confstring}, the string action is induced by an
antisymmetric tensor field. This action realizes explicitly
the necessary zig-zag invariance of confining strings
 \cite{zigzag}. It can be
derived without extra assumptions
\cite{dqt} for the confining phase of compact $U$(1) gauge
theories \cite{polbook}. The resulting strings
are dual versions of the magnetic strings
of the Abelian Higgs model \cite{kl,ahm}.
Moreover, the action is a
specific form
of an earlier-proposed class of actions
which would generate
surfaces with non-trivial elastic properties \cite{dyn}.

The characteristic of the effective string action obtained by
integrating out the tensor field is a non-local interaction
with negative stiffness between world-sheet elements. Such an interaction
has been shown \cite{chervyakov} to
produce the correct high-temperature behavior of confining strings known
from large-$N$ QCD calculations \cite{hit}. Moreover, it has been
proposed \cite{dt} as a mechanism leading to smooth strings, with
``long-range" orientational correlations, a fact confirmed by recent
numerical simulations \cite{polikarpov}.

In this letter we report on further progress in this field by
considering the simplest possible model in this class of actions,
obtained by adding to the rigid string the next-order gradient term.

\noindent
{\bf 2.} The new model is defined in euclidean space
by the action
\begin{equation}
S = \int d^2{\xi } \sqrt{g} \ g^{ab}
{\cal D}_a x_{\mu }\! \left( T - s{\cal D}^2 +
{1\over M^2} {\cal D}^4 \right)
\ {\cal D}_b x_{\mu } \ ,
\label{one}
\end{equation}
where ${\cal D}_a$ are covariant derivatives with
respect to the induced metric $g_{ab}=\partial_a x_{\mu }\partial _bx_{\mu }$
on the surface ${\bf x}\left( \xi _0, \xi_1\right) $ and we have used
units $c=1$, $\hbar=1$. Here, the bracket has to be considered as
representing the first few terms
in the expansion of the non-local interaction mediated by the original
antisymmetric tensor. The first term provides a bare surface tension
$2T$, while the second accounts for rigidity \cite{rigid}
with stiffness parameter $s$.
The last term can be written (up to surface terms)
as a combination of the fourth
power and the square of the gradient of the extrinsic curvature matrices,
with $M$ being a new mass scale. It thus suppresses world-sheet
configurations with rapidly changing extrinsic curvature; due to its
presence, the stiffness $s$ can be positive or negative, as is actually the
case \cite{dqt} when the stiffness is
generated dynamically by a tensor field in
four-dimensional space-time.
Note that the action (\ref{one}) is not
Weyl invariant, so that no conformal anomaly
appears.

We analyze the model (\ref{one}) in the
large-$D$ approximation
along the lines of Ref. \cite{largen}. To this end we introduce a
Lagrange multiplier matrix $L ^{ab}$
to enforce the constraint $g_{ab}=\partial _ax_{\mu }\partial_bx_{\mu }$,
extending the action (\ref{one}) to
\begin{equation}
 S+ \int d^2\xi \sqrt{g} \ \ L
^{ab} \left( \partial _a x_{\mu } \partial _bx_{\mu } - g_{ab} \right) \ .
\label{two}
\end{equation}
Then we parametrize the world-sheet in a Gauss map by
$x_{\mu } (\xi ) = \left( \xi _0, \xi _1, \phi ^i (\xi )
\right) $, $(i=2, \dots , D-1)$,
where $-\beta /2\le \xi_0 \le \beta /2$,
$-R/2 \le \xi _1 \le R/2$ and
$\phi ^i(\xi )$ describe the ($D$-2)
transverse fluctuations.
With the usual homogeneity and isotropy ansatz
$g_{ab}=\rho \ \delta_{ab} $,  $L ^{ab} =
L \ g^{ab}$
of infinite surfaces
($\beta ,R \to \infty $) at the saddle
point, we obtain
\begin{eqnarray}
S &&=2\int d^2\xi \ \left[ T +L
(1-\rho ) \right]
\nonumber \\
&&+ \int d^2\xi  \ \partial_a\phi ^i
\ V\left( T, s, M, L, {\cal D}^2
\right) \ \partial_a \phi ^i\ ,
\label{three}
\end{eqnarray}
where
\begin{eqnarray}
V\left( T, s, M, L, {\cal D}^2 \right) &&= T+L - s {\cal D}^2 +
{1\over M^2} {\cal D}^4 \ ,
\label{four}
\end{eqnarray}
Integrating over the transverse fluctuations, in the infinite
area limit, we get the effective action
\begin{eqnarray}
S^{\rm eff} &&= 2A_{\rm ext} \ \left[ T+L
(1-\rho ) \right]
\nonumber \\
&&+ A_{\rm ext} {{D-2}\over 8\pi^2 }\rho
\int d^2p\ {\rm ln}
\left[ p^2 V\left( T, s, M, L, p \right)
\right] \ ,
\label{five}
\end{eqnarray}
where $A_{\rm ext}=\beta R$ is the
extrinsic, physical, space-time area.
For large
$D$, the fluctuations of $L$ and $\rho $ are suppressed and these
variables take their ``classical values", determined by the two saddle-point
equations
\begin{equation}
0 = f\left( T, s, M, L \right) \ ,\quad 
\rho = {1\over f'\left( T, s, M , L \right) }\ ,
\label{six}
\end{equation}
where the prime denotes a derivative with respect to $L$ and
the ``saddle-function" $f$ is defined by
\begin{eqnarray}
f\left( T, s, M, L \right) &&\equiv L \\
\nonumber
&&- {{D-2}\over {8\pi }}
\int dp \ p\  {\rm ln}
\left[ p^2 V
\left( T, s, M, L, p \right) \right] .
\label{additional1}
\end{eqnarray}
Using (\ref{six}) in (\ref{five}) we get
$S^{\rm eff}= 2\left( T+L \right)
\ A_{\rm ext}$
showing that ${\cal T} =2\left( T+L \right)$
is the physical string tension.

The stability condition for the euclidean surfaces is that
$V\left( T, s, M, L, p \right)$ be positive for all $p^2 \ge 0$.
However, we shall require the same condition also for $p^2\le 0$,
so that strings propagating in Minkowski space-time are
not affected by the
propagating states of negative norm which plague rigid strings.
The stability condition becomes thus
$\sqrt{T+L} \ge |sM/2|$, which
allows us to introduce the real variables $R$ and $I$ defined by
\begin{equation}
R^2 \equiv {M\over 2} \sqrt{T+L}
+ {s M^2\over 4}\ ,\quad 
I^2 \equiv {M\over 2} \sqrt{T+L}
- {s M^2\over 4}\ .
\label{eight}
\end{equation}
In terms of these, the kernel $V$ can be written as
\begin{equation}
M^2 V\left( T, s, M, L, p \right) =
\left( R^2+I^2 \right) ^2 + 2 \left( R^2-I^2 \right) p^2
+ p^4 \ .
\label{newone}
\end{equation}

\noindent
{\bf 3.} In order to distinguish the various phases of our model we
compute two correlation functions. First, we
consider the orientational correlation function
$g_{ab}(\xi -\xi ') \equiv \langle \partial_a \phi ^i (\xi )
\ \partial_b \phi ^i (\xi ') \rangle $ for the normal components of
tangent vectors to the world-sheet. From (\ref{three}) this is given by
\begin{equation}
g_{ab}(\xi -\xi ') = {\delta _{ab} \over 8\pi ^2}
\ \int d^2p \ {1\over
V\left( T, s, M, L, p \right) }
\ \ {\rm e}^{i \sqrt{\rho } p (\xi -\xi ')} \ .
\label{ten}
\end{equation}
In terms of $R$ and $I$, the Fourier
components can be writtens as
\begin{equation}
{1\over V\left( T, s, M, L, p \right) } =
{M^2\over 2RI} \ {\rm Im} \ {1\over {p^2 + (R-iI)^2}} \ ,
\label{eleven}
\end{equation}
from where we obtain the following exact result for the diagonal
elements $g  \equiv g_{aa}$ of (\ref{ten}):
\begin{equation}
4 \pi  g (d) =
{M^2\over 2RI}
\ {\rm Im}\ K_0\left( \left( R-iI \right)
\sqrt{\rho} d \right) \ ,
\label{twelve}
\end{equation}
where $d\equiv |\xi -\xi'|$ and $K_0$ is a
Bessel function of imaginary argument
\cite{gr}.

Secondly, we compute the scaling law of
the distance $d_E$ in embedding space
between two points on the
world-sheet  when changing its projection $d$
on the reference plane. The exact relation between
the two lengths is
\begin{equation}
d_E^2=d^2 + \sum_i \langle |\phi ^i (\xi) - \phi ^i (\xi ')
|^2\rangle \ .
\label{newtwo}
\end{equation}
With a computation analogous to the one leading to (\ref{twelve})
we obtain the following behaviour:
\begin{eqnarray} \nonumber
d_E^2= \cases{\displaystyle{
\left( R^2+I^2 \right) \over 16 \pi {\cal T} RI}
\ {\rm arctan} (I/R) \ \alpha d^2, &$d^2 \ll
{1\over \alpha}$ \ ,\cr\cr
\displaystyle
{1\over 4\pi {\cal T}} \ {\rm ln}\left( \alpha
d^2 /4 \right) +{\rm const.}, 
&${1\over \alpha}\ll d^2
\ll {1\over 2\pi {\cal T}}$\ ,\cr \cr
\displaystyle
d^2, &$d^2\gg {1\over 2\pi {\cal T}}$\ ,%\cr
}
\nonumber
\end{eqnarray}
with $\alpha \equiv \left( R^2+I^2 \right) \rho $.

These results show that the model has three possible phases.
The first one is realized when there are no solutions to the saddle-point
equations in the allowed range of parameters,
which means that, for this choice of parameters, there exist no
homogeneous, isotropic surfaces. This is realized, as we shall see, for $R$
very small, when the spectrum of transverse fluctuations develops an instability
at a finite value $p=\sqrt{I^2-R^2}$.
In this phase (I) the surfaces will form inhomogeneous
structures \cite{david}.

If a solution to the saddle-point equations exists, two situations can
be realized. For large positive stiffness $s$ we have
$R\gg I$, the asymptotic region is $d
\gg 1/R\sqrt{\rho}$ and $I$ can be neglected.
In this region we have
\begin{equation}
g(d) \propto {1\over \sqrt{R\sqrt{\rho}d}} \ {\rm e}^
{-R\sqrt{\rho}d} \ ,
\label{thirteen}
\end{equation}
exhibiting short-range orientational order.
For short distances $d \ll 1/R\sqrt{\rho}$,
world-sheets behave as two-dimensional
objects.
If $\rho$ becomes large we have a region $1/R\sqrt{\rho} \ll
d\ll 1/\sqrt{2\pi {\cal T}}$ in which $d_E$ scales logarithmically with
$d$ and distances along the surface become large. The transition to this
regime happens on the scale of the persistence length
$d_E^{\rm PL}=1/\sqrt{{\cal T}}$.
Above this scale world-sheets are crumpled,
with no orientational correlations (if the tension is not large enough
to dominate over the entire surface,
causing $\rho \simeq 1$). This phase (R)
corresponds to the behaviour of the familiar rigid strings \cite{largen}.

For large negative stiffness $s$, in
contrast, we have
$I\gg R$, the asymptotic region is $d
\gg 1/I\sqrt{\rho}$, and $R$ can be neglected in $K_0$ for $d
\ll 1/R\sqrt{\rho} $. In this region we have
\begin{equation}
8 g(d)= {M^2\over 2RI}
\ J_0 \left( I\sqrt{\rho}d \right) \ ,
\label{fourteen}
\end{equation}
with $1/R\sqrt{\rho} $ playing the role of
an infrared cutoff for the oscillations
on the scale $1/I\sqrt{\rho}$ over which the Bessel function
$J_0$ varies. In this case $d_E^O=\sqrt{I/R{\cal T}}$ is the
scale on which world-sheets form oscillating
structures characterized
by the ``long-range antiferromagnetic" orientational
correlations (\ref{fourteen}). Crumpling takes place only if
$1/R\sqrt{\rho} \ll 1/\sqrt{2\pi {\cal T}}$ and the corresponding
persistence length is
$d_E^{\rm PL}=(1/\sqrt{{\cal T}})
\sqrt{ (\pi I/2R)+{\rm ln} \left( I^2/4R^2\right) } $,
which is much larger than $1/\sqrt{{\cal T}}$.
Otherwise, the oscillating superstructure,
which we shall call the {\it hyperfine
structure} of strings,
goes over directly into the tension dominated region.
Such world-sheets constitute a liquid
version of the ``egg carton" surfaces (LEC)
of the biomembrane literature
\cite{eggcarton}.
In this case $d_E$
scales logarithmically with $d$ for $1/I\sqrt{\rho} \ll d
\ll 1/\sqrt{2\pi {\cal T}}$, and $\rho $ is large not because
of crumpling but because of the many ``hills and valleys" on the
liquid egg carton; indeed there are strong orientational correlations in
this region. For very small $R$ one can thus obtain long strings
with an oscillating superstructure, a mechanism proposed
in \cite{dt}.
For symmetry reasons, the transition
from the rigid to the liquid egg carton
phase occurs on the line $R=I$, where $s=0$. On this line
the spectrum of transverse fluctuations,
$E\left( T, s, M, L, p\right) = p^2\ V\left( T, s, M, L, p\right)$
acquires a roton-like minimum, as shown schematically in Fig. 1.

\begin{figure}
\centerline{\epsfysize=5.5cm\epsfbox{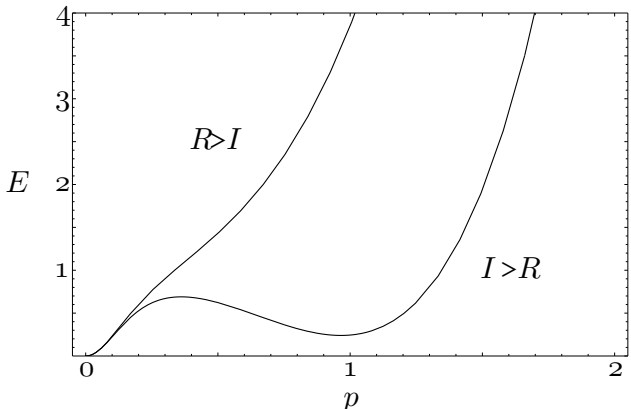}}
\caption{The roton-like minimum in the spectrum of transverse fluctuations
for $I>R$.}
\end{figure}

In order to fully establish the phase diagram
of our model, we analyze
the saddle-point function $f\left( T, s, M, L \right)$
in (\ref{additional1}). To this end we must
prescribe a
regularization for the ultraviolet divergent
integral. We use dimensional regularization,
computing
the integral in $(2-\epsilon)$ dimensions.
For small $\epsilon$, this leads to
\begin{eqnarray}
f\left( T, s, M, L\right) &&= L + {1\over 16 \pi ^3} \left( R^2-I^2
\right) {\rm ln} {R^2+I^2\over \Lambda ^2} \\
\nonumber
&& - {1\over 8 \pi ^3} RI \left( {\pi \over 2} + {\rm arctan}
{I^2-R^2 \over 2RI} \right) \ ,
\label{newfour}
\end{eqnarray}
where $\Lambda \equiv \mu \ {\rm exp} (2/\epsilon)$ and $\mu $ is a
reference scale which must be introduced for dimensional reasons.
The scale $\Lambda $ plays the role of an
ultraviolet cutoff,
diverging for $\epsilon \to 0$.

The saddle-point function above
is best studied by introducing the dimensionless
couplings $t\equiv T/\Lambda ^2$, $m\equiv M/\Lambda $ and
$l\equiv L/\Lambda ^2$.
A numerical analysis of the solutions of the saddle-point equation
$f\left( t, s, m, l\right) =0$ leads to
the phase diagram represented in Fig. 2
for $m=100$.

\begin{figure}
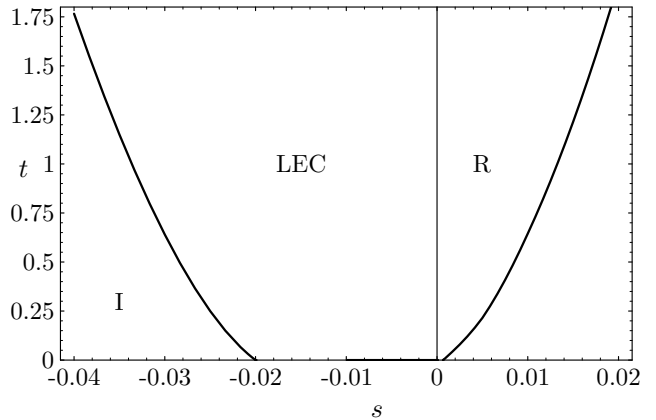

%\centerline{\epsfysize=5.5cm\epsfbox{quarticb.eps}}
~\\[-6mm]
\input  quarticb.tps
~\\
\caption[]{The three phases of our model as a function of the parameters $s$
and $t$ for $m=100$: an inhomogeneous (I) phase,
a rigid (R) phase and the new
liquid egg carton (LEC) phase.}
\end{figure}

On the separatrix between LEC and I we
have $\rho = \infty$, showing that
world-sheets crumple for large negative
stiffness and too small bare tension. The
same happens for positive stiffnesses up to
$s\simeq 0.002$. For larger positive
stiffnesses the graph represents the lower
boundary $I=0$ of the stability regime.
Notice that world-sheets never crumple for
small enough stiffnesses (negative and
positive). We therefore study in detail
the case $s=0$. In this case,
the saddle-point equations can be solved analytically
since $R=I$. We get
\begin{eqnarray}
l &&= {m^2 c^2\over 2}\ \left(
1+\sqrt{1+{4t\over m^2 c^2}} \right) \ ,\\
\rho &&= \left( 1-{m c\over 2\sqrt{t+l}} \right) ^{-1}\ ,
\label{newfive}
\end{eqnarray}
where $c\equiv 1/32 \pi ^2$. This shows that the point ($t^*=0$,
$s^*=0$, $m^*=0$) constitutes
an infrared-stable fixed-point with vanishing
physical string tension ${\cal T}$. This point is characterized by
long-range correlations
$g(d) = 2\pi ^2/a$,
with a constant $a$, and by the scaling law
\begin{equation}
d_E^2 = {\pi ^2\over a}\ \rho ^* \ d^2 \ , \quad
\rho ^* \equiv \left( 1-{1\over 2a} \right) ^{-1} \ ,
\label{newseven}
\end{equation}
which shows that the Hausdorff dimension of world sheets is
$D_H=2$. For $s=0$, the constant $a$ can be computed analytically:
\begin{equation}
a^2 = \lim _{{t\to 0}\atop {m\to 0}}
{{1+(2t/m^2 c^2) + \sqrt{1+ 4t/m^2 c^2 }}\over 2}\ ,
\label{neweight}
\end{equation}
from which we recognize that $1\le \rho ^* \le 2$.

At the infrared fixed-point we can remove the cutoff. The
renormalization of the model is easily obtained by noting that
the effective action for transverse fluctuations to quadratic
order decouples from other modes and is identical with the
second term in (\ref{three}) with ${\cal D}^2=\nabla ^2/\rho$
and $\rho $ taking its saddle-point value.
(see also \cite{largen}). From
here we identify the physical tension, stiffness and mass as
\begin{equation}
{\cal T} = \Lambda ^2 (t+l)\ ,\quad
{\cal S} = {s\over \rho}\ ,\quad
{\cal M} = \Lambda \ m \rho\ .
\label{newnine}
\end{equation}
For $s=0$, we can compute analytically the corresponding
$\gamma$-functions:
\begin{eqnarray}
\gamma_t &&\equiv -\Lambda {d\over d\Lambda} {\rm ln}\ t = 2+
O\left( {t\over m^2} \right) \ ,\\
\gamma _m  &&\equiv -\Lambda {d\over d\Lambda} {\rm ln}\ m = 1+
O\left( {t^2\over m^4} \right) \ .
\label{newten}
\end{eqnarray}

\noindent
{\bf 4.} In summary, the vicinity of the infrared fixed-point defines a new
theory of smooth strings for which the range of the orientational
correlations in embedding space
is always of the same order or bigger than the
length scale $1/\sqrt{{\cal T}}$
 associated with the tension. The
naively irrelevant term ${\cal D}^4/M^2$ in
(\ref{one}) becomes relevant in the
large-$D$ approximation since it generates a
string tension proportional to $M^2$ which
takes over the control of the fluctuations
after the orientational
correlations die off.
Note moreover, that it is exactly this new quartic
term which guarantees that the spectrum $p^2 V\left(
T, s, M, L, p\right)$ has no other pole than $p=0$,
contrary to the rigid string, which necessarily has
a ghost pole at $p^2= -T/s$.

\end{document}